# Development of high-power and ultra-high-vacuum waveguide coupler working at C/X band

He Xiang(贺祥), Zhao Feng-Li(赵风利), Wang Xiang-Jian(王湘鉴)

Institute of High Energy Physics, Chinese Academy of Sciences, Beijing 100049, China

**Abstract** Waveguide directional couplers working at 5.712/11.9924 GHz are developed. Even holes symmetrical to the structure are drilled along the central line of the narrow-wall of waveguide, which are used to couple the electromagnetic power from the main-waveguide to the sub-waveguide. The final prototypes have got satisfactory performances of high-power, ultra-high-vacuum and high-directivity. The microwave measurement results are also qualified.

**Key words**    high-power, ultra-high-vacuum, high-directivity, C-band, X-band, waveguide coupler.

**PACS**      Accelerators, 29.20.-c

## 1 Introduction

As one of the most widely used microwave components in many microwave systems, directional coupler is used to divide the input microwave power according to a given ratio. Different from using in microwave circuits or the radars, in the accelerator system[1][2], the directional couplers are mainly used for watching the power continuously or for chain protection together with a power meter and a control box.

Unlike commonly used in S-band, prototypes of new C/X band waveguide directional couplers with central frequency of 5.712/11.9924 GHz are developed, which are used in CERN's linear accelerator and PSI's free electron laser (FEL) respectively. The prototypes have got a vacuum leakage rate of less than $2\times10^{-10}$ Torr·L/s (ultra-high-vacuum), a directivity of more than 29 dB (high-directivity), and a very good high-power performance.

## 2 Structure

The structure (also the simulation model in CST 2010) of the directional couplers are shown in Fig.1. Six holes of the C-band coupler (four holes for X-band coupler) are drilled along the central line of the narrow-wall of the waveguide. The two ends of the main-waveguide are welded with flanges while two shorting walls are used at the two ends of the sub-waveguide, and two N-type joints for C-band coupler (SMA joints for X-band coupler) are welded on the wide-wall of the sub-waveguide to pick the microwave power out.

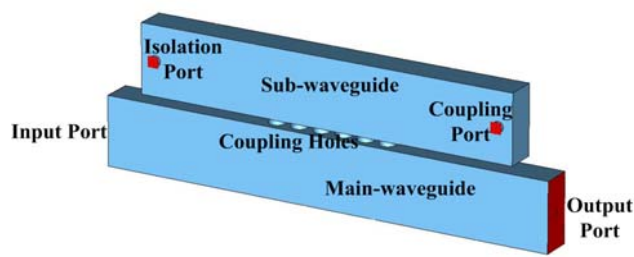

(a) C-band, Six coupling holes

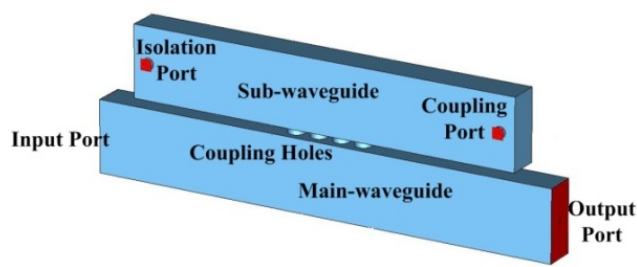

(b) X-band, Four coupling holes

Fig.1    Structure (simulation model) of the C/X band waveguide directional coupler (3D)

The microwave power comes in from the Input Port and very little power is coupled to the sub-waveguide and picked out by the joints finally.

## 3 Simulation result

The most important parameters of the couplers are the distance between the coupling holes, the diameter of each coupling hole, the distance between the central point of the SMA joint and the shorting wall, the insertion depth of the

SMA joints.

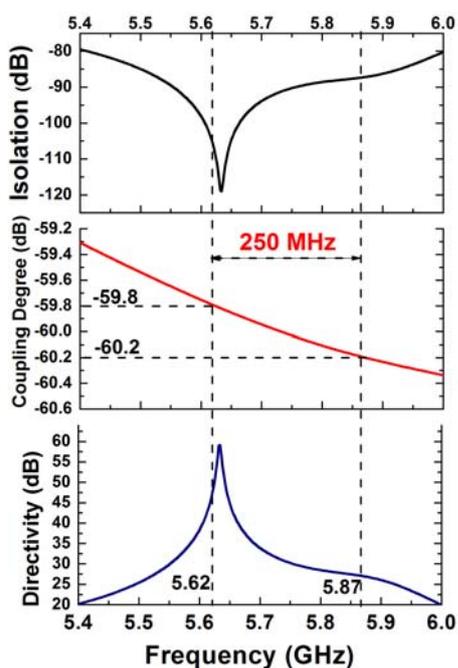

Fig. 2 Frequency response of Directivity, Coupling Degree and Isolation for C-band coupler

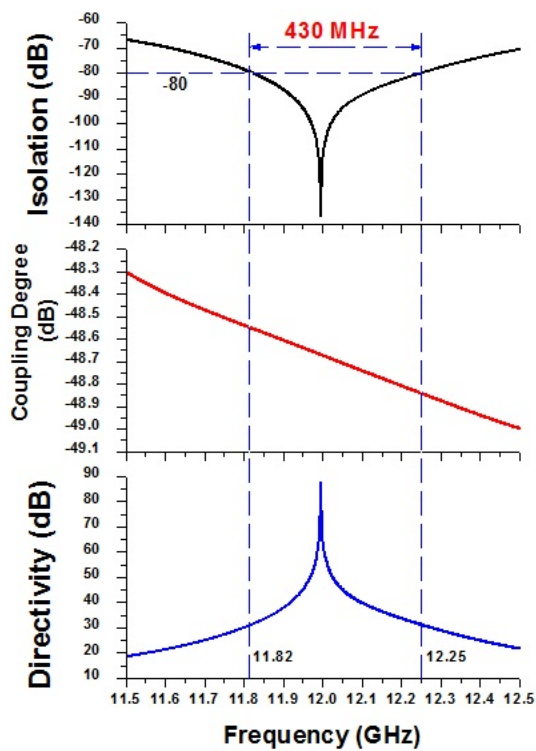

Fig. 3 Frequency response of Directivity, Coupling Degree and Isolation for X-band coupler

After the simulation and optimization, the developed directional couplers have got Directivity of more than 25/31 dB within a bandwidth of 250/430 MHz (5.62~5.87 GHz/ 11.82~12.25 GHz) while maintaining a VSWR less than 1.04/1.04 as well as a variation of the Coupling Degree less than 0.4/0.4 dB (-59.8~-60.2 dB/ -48.5~-48.9 dB) at the same time in simulation.

The frequency response of the Directivity of the C/X band couplers are shown in Fig. 2 and 3 respectively.

## 4  Measurement result

The prototypes of C/X band couplers are shown in Fig. 4, and the compare between the simulation and measurement results of C/X band couplers at their central frequencies are shown in Table 1 and 2 respectively. Note that the A/B in tables mean that the results are obtained when the microwave power coming inside from the left (Input port)/right (Output port) port (see in Fig. 1).

The C-band waveguide directional coupler has been accepted by PSI's FEL for meeting their requirements as: In bandwidth of 5712±3 MHz, VSWR less than 1.04, Coupling Degree within -60±0.2 dB, Directivity more than 25 dB at the same time. And the high power test of the C-band coupler has been performed in the PSI C-band test facility TRFCB01 from the end of March to the beginning of May in 2014. The schematic diagram of the test is shown in Fig. 5. The Conditioning took 4 weeks, during which the pulse length was increased from 100 ns to 3000 ns by steps of 100 ns when keeping the vacuum pressure always below $1\times10^{-7}$ mbar. After that, the test at constant power took 5 weeks, the power from klystron was set as rectangular pulse with pulse width of 3 μs and peak power of 40 MW (average power 12 kW for repetition rate of 100Hz), the vacuum pressure was always below $1\times10^{-8}$ mbar.

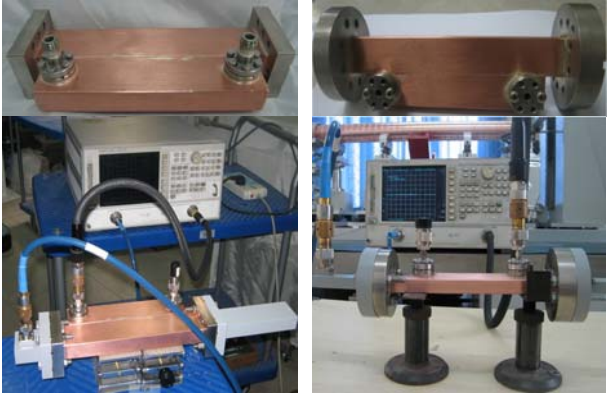

(a) C-band coupler    (b) X-band coupler
Fig. 4 Prototype and testing of the waveguide directional couplers

Table 1. Compare of simulation and measurement results of C-band coupler at 5.712 GHz

|  | Simulation | Measurement |
| --- | --- | --- |
| VSWR | 1.01/1.01 | 1.02/1.02 |
| Insertion Loss (dB) | -0.1/-0.1 | -0.5/-0.4 |
| Coupling Degree (dB) | -60.0/-60.0 | -58.5/-59.4 |
| Isolation (dB) | -93/-93 | -90/-92 |
| Directivity (dB) | 33/33 | 32/33 |

*Left port input/ Right port input

Table 2. Compare of simulation and measurement results of X-band coupler at 11.9924 GHz

|  | Simulation | Measurement |
| --- | --- | --- |
| VSWR | 1.04/1.04 | 1.07/1.04 |
| Insertion Loss (dB) | -0.002/-0.002 | -0.1/-0.1 |
| Coupling Degree (dB) | -48.7/-48.7 | -47.0/-47.4 |
| Isolation (dB) | -117/-117 | -79/-76 |
| Directivity (dB) | 68/68 | 32/29 |

*Left port input/ Right port input

From Table 1, it can be seen that the simulated and measured results match well, which means the C-band coupler a successful one. However, From Table 2, the measured results of X-band coupler is acceptable but worse than the simulated results. For the very strict requirement of accuracy of the fabrication in X-band, even a very small fabrication error will result in a very large deviation from the simulation results.

The prototypes have got a vacuum leakage rate of less than $2\times10^{-10}$ Torr·L/s and the C-band coupler has got a vacuum value better than $1\times10^{-8}$ mbar ($7.5\times10^{-9}$ Torr) during high power test, so the prototypes have got a good ultra-high-vacuum performance. Furthermore, the results of the high power and the microwave tests showed that the prototypes are of good high-power and high-directivity (Directivity of more than 32/29 dB for C/X band couplers respectively) performances as well.

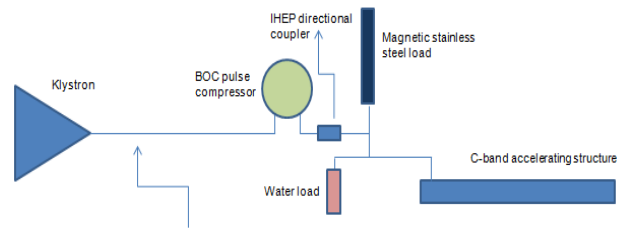

Fig. 5 The schematic diagram of the high power test in PSI

## 5  Conclusion

The C/X band waveguide directional couplers developed have got advantages such as high-power, ultra-high-vacuum, high-directivity, stable Coupling Degree, wide Bandwidth, etc. The high power test of the C-band coupler has been done by PSI, the result is satisfactory. However, for the X-band directional coupler, even a very small fabrication error may cause a very large deviation between the measured and simulated results because of the high accurate requirement of fabrication. So it is very important to control the fabrication error during the future processing.

# 6 Acknowledgement

I want to give my best thanks to Dr. Riccardo Zennaro of PSI who performed the high power test of C-band coupler and gave data to me. And I am also grateful to Dr. Germana Riddone of CERN who gave me lots of important information and help during the development of the X-band coupler.